\documentclass[twocolumn,showpacs,preprintnumbers,amsmath,amssymb]{revtex4}

\usepackage{graphicx}
\usepackage{dcolumn}
\usepackage{bm}

\begin{document}


\title {Symplectic structure of post-Newtonian Hamiltonian for spinning compact binaries} 

\author{Xin Wu}
\affiliation{Department of Physics,
Nanchang University, Nanchang 330031, China}

\author{Yi Xie}
\email{yixie@nju.edu.cn}
\affiliation{Astronomy Department, Nanjing University, Nanjing 210093, China\\
Department of Physics $\&$
Astronomy, University of Missouri-Columbia, Columbia, Missouri
65211, USA}


\begin{abstract}

The phase space of a Hamiltonian system is symplectic. However,
the post-Newtonian Hamiltonian formulation of spinning compact
binaries in existing publications does not have this property,
when position, momentum and spin variables $[\mbox{\boldmath$X$},
\mbox{\boldmath$P$}, \mbox{\boldmath$S$}_1,
\mbox{\boldmath$S$}_2]$ compose its phase space. This may give a
convenient application of perturbation theory to the derivation of
the post-Newtonian formulation, but also makes classic theories of
a symplectic Hamiltonian system be a serious obstacle in
application, especially in diagnosing integrability and
nonintegrability from a dynamical system theory perspective. To
completely understand the dynamical characteristic of the
integrability or nonintegrability for the binary system, we
construct a set of conjugate spin variables and reexpress the spin
Hamiltonian part so as to make the complete Hamiltonian
formulation symplectic. As a result, it is directly shown with the
least number of independent isolating integrals that a
conservative Hamiltonian compact binary system with both one spin
and the pure orbital part to any post-Newtonian order is typically
integrable and not chaotic. And a conservative binary system
consisting of two spins restricted to the leading order spin-orbit
interaction and the pure orbital part at all post-Newtonian orders
is also integrable, independently on the mass ratio. For all other
various spinning cases, the onset of chaos is possible.

\end{abstract}


\maketitle

\section{Introduction}

Gravitational waves from coalescing spinning compact binaries,
made of neutron stars and/or black holes, are important sources
for ground-based and future space-borne detectors. Since a
successful detection requires theoretical gravitational-wave
templates matched with experimental data, the dynamics of two
spinning compact bodies has recently been a hot issue in
post-Newtonian (PN) celestial mechanics. As a current breakthrough
in this field, there are several different methods for deriving
the equations of motion of two point-like particles up to 2PN
order [1,2] and 3PN order [3,4], and even up to higher-order PN
approximations in general relativity [5]. One of these methods
refers to the PN Lagrangian formulation, giving the orbits of
black hole pairs in harmonic coordinates and in a general frame
[4]. Another deals with the PN Hamiltonian formulation, describing
the motion of two compact bodies in Arnowitt- Deser-Misner (ADM)
coordinates and in the center-of-mass frame [3]. These two
formulations have been proved to be approximately but not exactly
equivalent [6,7]. It should be noted that the so-called physical
\emph{equivalence} between the two approaches is only based on a
certain PN order accuracy, namely, has a small difference in the
PN order approximation. For instance, conserved quantities of
motion (if they exist) for the former are generally accurate to
the PN order level, while they are rigorously invariant for the
latter.

However, there may be a great difference between the two
formulations from a dynamical point of view. In the case of a
comparable mass binary system with only one spinning body, the 2PN
Hamiltonian dynamics shows no chaos due to the integrability of
the system [8,9], but the 2PN Lagrangian dynamics was identified
to be chaotic by the method of fractal basin boundaries built on
the unstable, fractal set of periodic orbits [10,11]. These
results are still correct when the two approaches give place to
the 2PN Hamiltonian formulation of two equal-mass compact objects
with two spins having the spin effects restricted to the leading
order spin-orbit interaction [8,9] and its corresponding 2PN
Lagrangian formulation [12], respectively. As a little attention
to deserve, these distinct results seem to be explicit conflict if
the difference between the two approaches is neglected. In this
sense, it is natural to initially yield some doubt regarding the
results. In fact the related arguments had never stopped until
Levin [11] pointed out that these results are stemmed from
different approximations to the same physical problem. Of course,
the use of different indicators of chaos is also an important
source leading to these arguments. At the beginning, the presence
of chaos in the conservative 2PN Lagrangian dynamics of spinning
compact binaries was confirmed by means of the fractal basin
boundary method [13]. This implies that there are unpredictable
gravitational waveforms during the inspiral. At once, the claim
was questioned in Ref. [14] that calculates Lyapunov exponents as
the divergence rate of nearby trajectories and finds no positive
but zero Lyapunov exponents in all cases tested. The analysis on
the chaos ruled out was strongly criticized in Refs. [12,15],
where positive Lyapunov exponents are  still obtained and the
reason for the false Lyapunov exponents appeared in Ref. [14] is
attributed to continually rescaling the shadow trajectory. In
spite of this, we showed in a previous article [16] that no
space-time coordinate redefinition ambiguity mentioned in [17] but
a slightly different computational treatment of Lyapunov exponents
is an exact source for these different results between Ref. [14]
and Ref. [15]. The Lyapunov exponents in Ref. [14] are determined
by the limit method for the computation of the stabilizing limit
values as reliable values of Lyapunov exponents, while they are in
Refs. [12,15] given by the fit method taking the slopes of the fit
line about the natural logarithm of the divergence rate of nearby
trajectories vs time as values of Lyapunov exponents. Clearly, the
limit method becomes more difficult to detect chaos from order
than the fit method when integration time is not long enough.
Further, we argued the onset of the chaotic behavior of a pair of
comparable mass black holes having one spin or two spins for the
2PN Lagrangian formulation by the invariant fast Lyapunov
indicators of two nearby trajectories proposed in Ref. [18],
viewed as a more sensitive tool to find chaos. Other reference
[19] also supported this fact with aid of the frequency map
analysis. As an exceptional case, the radial motion of spinning
compact binaries in the Lagrangian formulation with contributions
from the spins, mass quadrupole and magnetic dipole moments is
explicitly integrated [20]. In addition, chaos in the conservative
2PN or 3PN Hamiltonian approach of compact binaries having two
spins can be seen from the paper [21] of Hartl and Buonanno who
adopted the fit method to calculate the Lyapunov exponents. As a
point to emphasize, although both Ref. [10] and Ref. [21] admit
the existence of chaos in the conservative 2PN Lagrangian or
Hamiltonian dynamics of compact binaries with two spins, there are
different opinions with respect to the dependence of chaos on
dynamical parameters and initial conditions. For example, it was
said in Ref. [10] that chaos becomes strong for the spins
perpendicular to the orbital angular momentum, but it was reported
in Ref. [21] that chaos is greatly possible to occur when initial
spin vectors are nearly antialigned with the orbital angular
momentum for the $(10+10)M_{\odot}$ configuration ($M_{\odot}$
being mass of the Sun). The reason for the discrepancy was
explained in our another work [22]. It was shown that no single
physical parameter or initial condition but a complicated
combination of all parameters and initial conditions affects the
transition to chaos. As concluded in [22], one should distinguish
these distinct results on chaos and order of spinning compact
binaries in some references according to different approximations
to this physical model, methods finding chaos, dynamical
parameters and initial conditions.

It should be noticed that the above results (except those in Refs.
[8,9]) associated to the dynamics of order and chaos are all from
numerical investigations. In principle, numerical investigations,
which do closely depend on numerical integrators as well as
indicators of chaos, dynamical parameters and initial conditions,
are only a check of the local dynamics but not a check of the
global dynamics. The so-called global structure of phase space
scanned by the fractal basin boundary method [11] or the fast
Lyapunov indicators [22] is still based on some specific dynamical
parameters and initial conditions. In this case, it is regarded to
as a partial but not thorough check. Additionally, although the
method of finding parametric solutions to the Hamiltonian dynamics
used in Refs. [8,9] is thought as an analytical method that can
study the global dynamics, it has the limitation of application.
In fact, a better and more rigorous method is to use the least
number of independent constants of motion as a criterion for the
prediction of the integrability or the nonintegrability hiding
possible and potential chaos (the relationship between the
integrability and the least number of independent constants will
be introduced in Section II). Especially for these conservative
Hamiltonian formulations of spinning compact binaries in which the
constants of motion are exactly conserved, the method should work
well without question. Unfortunately, a problem lies in that the
phase space made of position, momentum and spin variables
$[\mbox{\boldmath$X$}, \mbox{\boldmath$P$}, \mbox{\boldmath$S$}_1,
\mbox{\boldmath$S$}_2]$ of these Hamiltonians is not completely
symplectic in known literature [3, 23-27]. Although this plays an
important role in providing the PN Hamiltonian formulation by the
convenient application of perturbation theory, it has an obvious
disadvantage that many Hamiltonian system properties have no way
to be applied for these systems. For example, a closed
nondegenerate differential 2-form, i.e., the so-called standard
symplectic structure on a manifold [28] cannot be defined clearly.
In particular, the relationship between the integrability and the
least number of independent constants cannot be understood
definitely. In view of the need of both a complete Hamiltonian
theory and a dynamical system theory, the main motivation of the
present paper is to design a group of new spin variables to
rewrite the spin Hamiltonian part of the conservative PN
Hamiltonian approximation for spinning compact binaries and to
make the phase space of the whole Hamiltonian system have the
symplectic structure so that we can apply the least number of
independent constants to judge the integrability or the
nonintegrability of the symplectic binary system and can further
provide some theoretical insight into the global dynamics.

The rest of this paper is organized as follows. At first, we
present a set of conjugate spin variables with the symplectic
structure in Section II. Then, several advantages of using them
are listed in Section III. Finally, Section IV summarizes the main
results.

\section{Construction of conjugate spin coordinates}

In this section, let us introduce the conservative PN Hamiltonian
formulation of spinning compact binaries, where the pure orbital
(nonspinning) part is accurate to 3PN order, and the spin part
arrives at the 4PN order approximation. Meanwhile, the evolution
equations of state variables and conserved quantities of motion in
the system are given. In addition, a method finding conjugate spin
variables with the symplectic structure is presented.

\subsection{Conserved quantities in the PN Hamiltonian formulation
 of spinning compact binaries}

The conservative Hamiltonian of spinning compact binaries is
\begin{equation}
H(\mbox{\boldmath$X$}, \mbox{\boldmath$P$}, \mbox{\boldmath$S_1$},
\mbox{\boldmath$S_2$}) = H_{O}(\mbox{\boldmath$X$},
\mbox{\boldmath$P$}) +  H_{S}(\mbox{\boldmath$X$},
\mbox{\boldmath$P$}, \mbox{\boldmath$S$}_1, \mbox{\boldmath$S$}_2)
\end{equation}
with the pure orbital part
\begin{equation}
H_{O} = H_{O,N} + H_{O,1PN} + H_{O,2PN} + H_{O,3PN}
\end{equation}
and the spin part consisting of spin-orbit (SO) coupling,
spin-spin (S$^{2}$) coupling and higher-order spin effects
\begin{eqnarray}
H_{S} &=& H_{SO,1.5PN} + H_{SO,2.5PN} + H_{S^{2},2PN}  \nonumber \\
& & + H_{S^{2}P^{2}, 3PN} + H_{S^{3}P, 3.5PN} +  H_{S^{4}, 4PN}.
\end{eqnarray}
The notation $S^{2}P^{2}$ represents various possible coupling
terms with respect to quadric forms of momentum
$\mbox{\boldmath$P$}$ and those of spins $\mbox{\boldmath$S_1$}$
and $\mbox{\boldmath$S_2$}$. The notation $S^{3}P$ refers to
couplings between momentum $\mbox{\boldmath$P$}$ and cubic terms
of $\mbox{\boldmath$S_1$}$ and $\mbox{\boldmath$S_2$}$, and
$S^{4}$ stands for quartic terms of $\mbox{\boldmath$S_1$}$ and
$\mbox{\boldmath$S_2$}$. Ref. [24] has given $H_{O}$, $
H_{SO,1.5PN}$ and $H_{S^{2},2PN}$ in the center-of-mass frame.
$H_{SO,2.5PN}$ can be found in [25], and $H_{S^{4}, 4PN}$ is
calculated in [26]. In addition, $H_{S^{2}P^{2}, 3PN}$ and
$H_{S^{3}P, 3.5PN}$ are provided in [27]. Note that these
Hamiltonians are directly given in the general frame, but they are
easily changed into ones in the center-of-mass frame by means of
the relation between the two frames given in [4]. Position
$\mbox{\boldmath$X$}$ and momentum $\mbox{\boldmath$P$}$ are  a
set of canonical variables that satisfy the Hamiltonian equations
of motion
\begin{equation}
\frac{d\mbox{\boldmath$X$}}{dt}=\frac{\partial H}{\partial
\mbox{\boldmath$P$}}, \quad \quad
\frac{d\mbox{\boldmath$P$}}{dt}=-\frac{\partial H}{\partial
\mbox{\boldmath$X$}}.
\end{equation}
The spin-evolution equations read [23]
\begin{equation}
\frac{d\mbox{\boldmath$S$}_i}{dt}=\frac{\partial H_{S}}{\partial
\mbox{\boldmath$S$}_i}\times \mbox{\boldmath$S$}_i ~~(i=1,2).
\end{equation}

Besides the total energy (1), five conserved quantities for the
system are the constant magnitude of spin vectors
\begin{equation}
\mbox{\boldmath$S$}^{2}_i=(\chi_{i}m^{2}_{i})^{2} ~~(i=1,2),
\end{equation}
and the total angular momentum [23]
\begin{equation}
\mbox{\boldmath$J$}=\mbox{\boldmath$L$}+ \mbox{\boldmath$S$}_1 +
\mbox{\boldmath$S$}_2
\end{equation}
with the Newtonian-looking angular momentum $\mbox{\boldmath$L$}=
\mbox{\boldmath$X$} \times \mbox{\boldmath$P$}$. Here
dimensionless spin parameters $\chi_{i}\in [0, 1]$ are allowed for
physically accessible realistic black hole or neutron star spins,
and $ m_{i}$ denotes the mass of body $i$. In short, there are six
independent constants or integrals of motion \footnote{Strictly
speaking, a constant of motion and an integral of motion are two
different concepts. An apparent difference between them lies in
that the former is a function of coordinates, velocities (or
momenta) and time, while the latter is a function of the phase
space coordinates. Every integral is a constant of the motion, but
the converse may not be true. Here the constant considered is only
a function of coordinates and momenta, so they are both the
same.}.

Now there is a problem whether the system (1) having the 6
integrals of motion in the 12-dimensional space made of
$[\mbox{\boldmath$X$}, \mbox{\boldmath$P$}, \mbox{\boldmath$S$}_1,
\mbox{\boldmath$S$}_2]$ is integrable. In order to answer it, let
us recall the criterion of integrability of a Hamiltonian system.
One must know $2n$ first integrals so as to obtain the analytical
solutions of a system of $2n$ ordinary differential equations
[28]. But it is often sufficient to know only $n$ first integrals
for a \emph{canonical} system of differential equations, whose
phase space is symplectic. Precisely speaking, the criterion of
integrability is attributed to Liouville's theorem that an
autonomous Hamiltonian with $n$ degrees of freedom (i.e., with a
$2n$-dimensional phase space) is integrable if it has $n$
independent integrals in involution [28]. Saying this in another
way, a canonical Hamiltonian with $n$ degrees of freedom is
integrable if and only if there are $n$ independent isolating
integrals [29]. Strong Jeans theorem [30] implies that the $n$, as
the required least number of independent isolating integrals for
identifying the integrability, corresponds to the case of all
regular with incommensurable frequencies. It can be inferred from
this criterion that the number of isolating integrals for the
integrability of the system (1) should be at least 9 rather than 6
due to the use of the spin-evolution equations (5) unlike the
canonical equations (4), that is, the global phase space of the
system (1) being nonsymplectic. Thus the existence of the above 6
integrals does not sufficiently show the integrability of the
system (1). As mentioned in the Introduction, this is also checked
numerically in the work [21]. To form this symplectic structure,
we will construct new spin variables in place of the old ones.

\subsection{A transformation to conjugate spin variables}

Let the spin vectors be expressed in \emph{cylindrical-like
coordinates} $(\rho_{i}, \theta_{i}, \xi_{i})$ as
\begin{equation}
\mbox{\boldmath$S$}_i = \chi_{i}m^{2}_{i}
\mbox{\boldmath$\hat{S}$}_i
\end{equation}
with unit spin vectors
\begin{equation}
\mbox{\boldmath$\hat{S}$}_i=\left (
\begin{array}{lcrccc}
\rho_{i} \cos\theta_{i}  \\
\rho_{i}\sin\theta_{i} \\
k_{i} \xi_{i}
\end{array}
 \right ),
\end{equation}
where each $\rho_{i}$ depends on $\xi_{i}$ as follows
\begin{equation}
\rho_{i} = \sqrt{1-(k_{i}\xi_{i})^{2}}.
\end{equation}
In the above equation, two coefficients $k_i$ are what we shall
determine. In fact, Eq. (8) gives a transformation from the old
spin variables to the new ones in the form
\begin{equation}
\mbox{\boldmath$S$}_i: ~ (S_{i1}, S_{i2}, S_{i3})\rightarrow
(\theta_{i}, \xi_{i}),
\end{equation}
where subscript $j$ denotes the $j$th-component $S_{ij}$ of the
spin vector $\mbox{\boldmath$S$}_i$. That is to say, each spin
containing 3 Cartesian spin components is a function of the 2 new
spin variables, marked as $\mbox{\boldmath$S$}_i
=\mbox{\boldmath$S$}_i (\theta_{i}, \xi_{i})$.

Using the new spin variables $[\theta_{1}, \theta_{2}, \xi_{1},
\xi_{2}]$, namely inserting Eq. (8) into Eq. (3), we rewrite
$H_{S}$ as
\begin{equation}
\mathbb{H}_{S}(\mbox{\boldmath$X$}, \mbox{\boldmath$\theta$};
\mbox{\boldmath$P$}, \mbox{\boldmath$\xi$}) =
H_{S}[\mbox{\boldmath$X$}, \mbox{\boldmath$P$},
\mbox{\boldmath$S$}_1 (\theta_{1}, \xi_{1}), \mbox{\boldmath$S$}_2
(\theta_{2}, \xi_{2})]
\end{equation}
with $\mbox{\boldmath$\theta$}=(\theta_{1}, \theta_{2})$ and
$\mbox{\boldmath$\xi$}=(\xi_{1}, \xi_{2})$. Suppose that
$(\theta_{i}, \xi_{i})$ are canonical coordinates in the phase
space with symplectic structure, and then we have canonical spin
Hamiltonian equations
\begin{equation}
\frac{d\theta_{i}}{dt}=\frac{\partial \mathbb{H}_{S}}{\partial
\xi_{i}}, \quad \quad  \frac{d\xi_{i}}{dt}=-\frac{\partial
\mathbb{H}_{S}}{\partial \theta_{i}}.
\end{equation}
It is clear that this hypothesis is true if  Eq. (5) is equivalent
to Eq. (13). The details of derivation are described in the
following.

It is easy to obtain
\begin{eqnarray}
\frac{\partial \mathbb{H}_{S}}{\partial \xi_{i}} &=&
\sum^{3}_{j=1} \frac{\partial H_{S}}{\partial S_{ij}}
\frac{\partial S_{ij}}{\partial \xi_{i}} =
k_{i}\chi_{i}m^{2}_{i}\bigg[-\frac{k_{i}\xi_{i}} {\rho_{i}}
\nonumber \\
& & \times\bigg(\frac{\partial H_{S}}{\partial S_{i1}}\cos\theta_{i}+
\frac{\partial H_{S}}{\partial S_{i2}}\sin\theta_{i}\bigg)+
\frac{\partial H_{S}}{\partial S_{i3}}\bigg], \nonumber \\
\frac{\partial \mathbb{H}_{S}}{\partial \theta_{i}} &=&
\sum^{2}_{j=1} \frac{\partial H_{S}}{\partial S_{ij}}
\frac{\partial S_{ij}}{\partial \theta_{i}} \nonumber \\ &=&
-\chi_{i}m^{2}_{i}\rho_{i}
 \bigg(\frac{\partial H_{S}}{\partial
S_{i1}}\sin\theta_{i}- \frac{\partial H_{S}}{\partial
S_{i2}}\cos\theta_{i}\bigg). \nonumber
\end{eqnarray}
According to the transformation (8), we have
\begin{eqnarray}
\begin{array}{lllllll}
& \frac{d\mbox{\boldmath$S$}_i}{dt} = \chi_{i}m^{2}_{i}
\frac{d\mbox{\boldmath$\hat{S}$}_i}{dt} \nonumber \\
&= \chi_{i}m^{2}_{i} \left (
\begin{array}{lcrccc}
-\frac{k^{2}_{i}\xi_{i}} {\rho_{i}} \cos\theta_{i}
\frac{d\xi_{i}}{dt}- \rho_{i}\sin\theta_{i} \frac{d\theta_{i}}{dt}\\
-\frac{k^{2}_{i}\xi_{i}} {\rho_{i}} \sin\theta_{i}
\frac{d\xi_{i}}{dt}+ \rho_{i}\cos\theta_{i} \frac{d\theta_{i}}{dt} \\
~~~~~~~~~~~~~~~~ k_{i}\frac{d\xi_{i}}{dt}
\end{array}
 \right )    \\
&= \chi_{i}m^{2}_{i} \left (
\begin{array}{lcrccc}
\frac{k^{2}_{i}\xi_{i}} {\rho_{i}} \cos\theta_{i} \frac{\partial
\mathbb{H}_{S}}{\partial \theta_{i}}- \rho_{i}\sin\theta_{i}
\frac{\partial \mathbb{H}_{S}}{\partial
\xi_{i}}\\
\frac{k^{2}_{i}\xi_{i}} {\rho_{i}} \sin\theta_{i} \frac{\partial
\mathbb{H}_{S}}{\partial \theta_{i}}+ \rho_{i}\cos\theta_{i}
\frac{\partial \mathbb{H}_{S}}{\partial
\xi_{i}} \\
~~~~~~~~~~~~ -k_{i}\frac{\partial \mathbb{H}_{S}}{\partial
\theta_{i}}
\end{array}
 \right )     \\
&= k_{i}(\chi_{i}m^{2}_{i})^{2} \left (
\begin{array}{lcrccc}
k_{i}\xi_{i} \frac{\partial H_{S}}{\partial S_{i2}}-
\rho_{i}\sin\theta_{i} \frac{\partial H_{S}}{\partial
S_{i3}}\\
-k_{i}\xi_{i} \frac{\partial H_{S}}{\partial S_{i1}}+
\rho_{i}\cos\theta_{i} \frac{\partial H_{S}}{\partial
S_{i3}} \\
\rho_{i}(\frac{\partial H_{S}}{\partial S_{i1}}\sin\theta_{i} -
\frac{\partial H_{S}}{\partial S_{i2}}\cos\theta_{i})
\end{array}
 \right )  \\
&= k_{i}\chi_{i}m^{2}_{i} \left (
\begin{array}{lcrccc}
S_{i3} \frac{\partial H_{S}}{\partial S_{i2}}- S_{i2}
\frac{\partial H_{S}}{\partial
S_{i3}}\\
-S_{i3} \frac{\partial H_{S}}{\partial S_{i1}}+ S_{i1}
\frac{\partial H_{S}}{\partial
S_{i3}} \\
S_{i2}\frac{\partial H_{S}}{\partial S_{i1}} - S_{i1}
\frac{\partial H_{S}}{\partial S_{i2}}
\end{array}
 \right )    \\
&= k_{i}\chi_{i}m^{2}_{i}
\begin{array}{lcrccc}
\frac{\partial H_{S}}{\partial \mbox{\boldmath$S$}_i}\times
\mbox{\boldmath$S$}_i.
\end{array}
\end{array}
\end{eqnarray}
If we take $k_{i}=1/(\chi_{i}m^{2}_{i})$, the above equation just
agrees with Eq. (5). Inversely, in the similar way we can also
derive Eq. (13) from Eq. (5). Therefore, conjugate spin variables
$(\theta_{i}, \xi_{i})$ whose time evolutions are given by Eq.
(13) are what we want. The system (1) and the precession equations
(5) can be reexpressed as a new complete canonical formalism
\begin{equation}
\mathbb{H}(\mbox{\boldmath$X$}, \mbox{\boldmath$\theta$};
\mbox{\boldmath$P$}, \mbox{\boldmath$\xi$}) =
H[\mbox{\boldmath$X$}, \mbox{\boldmath$P$}, \mbox{\boldmath$S$}_1
( \theta_{1}, \xi_{1}), \mbox{\boldmath$S$}_2 (\theta_{2},
\xi_{2})]
\end{equation}
and the precession equations (13), respectively. This means that
there are only two independent new spin variables in the spin
precession equations (13) per compact body. Here we specify
$\chi_{i}\neq 0$. If one of $\chi_{1}$ and $\chi_{2}$ vanishes,
the other nonzero spin vector needs rewriting in the form (8). If
$\chi_{1}=\chi_{2}=0$, the pure orbital part itself is of the
canonical formalism. We also find that it is impossible to get
conjugate spin variables if the original spin vectors
$\mbox{\boldmath$S$}_i$ are expressed in \emph{spherical
coordinates}.

It should again be emphasized that the definition of the word
``canonical" mentioned above does completely coincide with one
given by the book entitled Classical Mechanics [31]. In other
words, two components $\theta_{i}$ and $\xi_{i}$ of each spin are
said to be canonical or conjugate variables if their time
evolutions can satisfy Eq. (13). In this case, the phase space of
the system (14) is completely symplectic. In a word, the canonical
or conjugate variables we called in this paper can equip the phase
space of the system (14) with a complete symplectic structure. As
an important illustration, the term ``canonical spin" appeared in
some references [32-34] means using canonical Dirac brackets
instead of the Poisson brackets when the equations of motion are
derived from that Hamiltonian. An explicit difference between
their spin variables and ours lies in that the former appears as a
spin tensor, while the latter relates to a two-dimensional vector.
Of course, the spin tensor can also be defined as a
three-dimensional spin vector like Eq. (4.26) of Ref. [33]. Still
the spin-evolution equations do resemble Eq. (5) rather than Eq.
(13). The facts have shown clearly that the meaning of the
canonical in these articles is not consistent with ours.

\section{Advantages of using the new spin variables}

It can easily be observed that the expression of $\mathbb{H}$ is
more complicated than that of $H$. This may explain why known
references use the old spin variables rather than the new ones to
derive the Hamiltonian formulations. Nevertheless, the use of the
new spin variables has more advantages from Hamiltonian dynamics.
We list following several main points.

(\emph{i}) Reduction of dimensionality. The use of the new spin
variables automatically satisfies the two constraints by Eq. (6)
such that a problem of 12-dimensional space is reduced to one of
10-dimensional phase space. That is to say, the 12 components of
$[\mbox{\boldmath$X$}, \mbox{\boldmath$P$}, \mbox{\boldmath$S$}_1,
\mbox{\boldmath$S$}_2]$ in Eq. (1) are changed into the 10
components of $[\mbox{\boldmath$X$}, \mbox{\boldmath$\theta$};
\mbox{\boldmath$P$}, \mbox{\boldmath$\xi$}]$ in Eq. (14). The
reduction of two variables can also be seen from the
transformation (11). It should be noted particularly that the
information on the constancy of the magnitude of the old spin
vectors does hide in the new variables. In fact, for any time the
new variables are always constrained by unit spin vectors (9) when
$\theta_{i}\in [0, 2\pi]$ and $\xi_{i}\in [-\chi_{i}m^{2}_{i},
\chi_{i}m^{2}_{i}]$.

(\emph{ii}) Symplectic geometry. The complete canonical formalism
(14) contains the symplectic structure expressed as
\begin{equation}
\mbox{\boldmath$\omega$}^{2}=\sum^{3}_{j=1} dX_{j}\wedge dP_{j} +
\sum^{2}_{i=1} d\theta_{i}\wedge d\xi_{i}.
\end{equation}
Then the volume form of the 10-dimensional phase space is defined
by the local coordinate representation
\begin{equation}
\mbox{\boldmath$\omega$}^{10}=\Pi^{3}_{j=1} dX_{j}\wedge dP_{j}
\wedge \Pi^{2}_{i=1} d\theta_{i}\wedge d\xi_{i}.
\end{equation}
The symbol $\wedge$ means wedge product. Meanwhile the integral
invariants of Poincar\'{e} exist. In short, the related theories
of symplectic geometry [28] are fully suitable for the completely
canonical Hamiltonian system, $\mathbb{H}$.

(\emph{iii}) Symplectic integration algorithms. A class of
important numerical schemes, called symplectic integrators
[35-37], can be used to  preserve both  the accuracy of essential
properties and the symplectic structure of the canonical system
$\mathbb{H}$. Unfortunately, the system is difficultly separated
into two integrable pieces such that explicit symplectic
integrators become useless. In spite of this, implicit symplectic
methods such as the implicit midpoint method [36] are always
efficient.

(\emph{iv}) Dynamics of compact binaries with one spinning body. A
conservative Hamiltonian binary system is an 8-dimensional
dynamical problem with the symplectic structure if only one body
spins. This symplectic system holds four constants of motion
including the total energy and the total angular momentum. This
sufficiently shows the integrability of the symplectic system,
\emph{regardless of} the PN order. Orbits are confined to a
4-dimensional torus. This is an extension to the result of Refs.
[8,9] that there is no chaos in the 2PN Hamiltonian formulation of
two compact objects when one body spins.

Perhaps someone casts doubt on the convincing of the result that
the conservative symplectic Hamiltonian for one spinning body at
any PN order to either the pure orbital part or the spin part is
completely integrable because the symplectic Hamiltonian is not
known at all PN orders. It just demonstrates the superior
properties of the symplectic Hamiltonian system. We emphasize
again that the result is always correct \emph{only if} the future
higher-order PN approximations are given in ensuring the existence
of these four integrals including the total energy and the total
angular momentum. Even the result ought to hold in the extreme
mass ratio limit at all PN orders. This seems to have an apparent
contradiction with Suzuki $\&$  Maeda' result that the dynamics is
chaotic at least for unphysical large values of the spin [38]. In
spite of the limit case, the dynamical model considered in this
paper and the system consisting of a Schwarzschild black hole and
a spinning test particle in Ref. [38] are still not equivalent.
Four typical differences between them are listed here. (a)
Mechanisms of dynamical approaches. The former is from PN
approximations, while the latter using the Papapetrou-Dixon
equations of motion is fully relativistic. (b) Symplectic
structure. The former belongs to a symplectic Hamiltonian system,
but the latter does not. (c) Dimensions of variables. For the
former the position, momentum and spin vectors are 3, 3 and 2
dimensions, respectively, while each of the position, momentum and
spin vectors has 4 dimensions for the latter. (d) Number of
integrals. In particular, the difference between them becomes
clearer by counting the number of their integrals. For the former
the four integrals are always present in the 8-dimensional phase
space with the symplectic structure, but for the latter one finds
the presence of five constraints including the relation for the
mass of the particle, the constant magnitude of the spin vector,
the spin supplementary condition, the energy of the particle and
the $z$ component of the total angular momentum. As stated above,
the five constraints are much less than the required least number
of independent isolating integrals for the integrability of the
12-dimensional nonsymplectic system. This fact is supported by the
result of Suzuki $\&$ Maeda. Recently, Ref. [39] also found
chaotic orbits for smaller spin values in the motion of spinning
test particles around a Schwarzschild field. Therefore, it can be
concluded that there is actually no explicit conflict between our
result and Suzuki $\&$ Maeda' result. In addition, as far as the
corresponding PN Lagrangian formulation associated with the PN
Hamiltonian of one spinning body is concerned, the so-called
constants except the constant magnitude of the spin are
approximately conserved at a certain PN order, that is, the
desired least number of integrals for the integrability is not
rigorously satisfied, so it is no surprise to see the onset of
chaos [10,11].

(\emph{v}) Dynamics of compact binaries having two spins. The four
conserved quantities involving the total energy and the total
angular momentum (7) do not sufficiently show that the symplectic
system $\mathbb{H}$ in the 10-dimensional phase space is
integrable. In fact, a fifth integral of motion is absent. Thus
the symplectic system is nonintegrable. From this point of view,
it is easy to understand the result of [21] that the 2PN
Hamiltonian binary system consisting of the leading order
spin-orbit integration and the spin-spin coupling is chaotic for
some specific dynamical parameters and initial conditions. On the
other hand, a conservative symplectic Hamiltonian spinning binary
system containing both the pure orbital part at all PN orders and
the leading order spin-orbit coupling has two additional conserved
quantities [23]:
\begin{equation}
\mbox{\boldmath$L$}\cdot \mbox{\boldmath$L$} = const, \quad \quad
\mbox{\boldmath$L$}\cdot \mbox{\boldmath$S$}_{eff} = const
\end{equation}
with $\mbox{\boldmath$S$}_{eff} =[2+3m_2/(2m_1)]
\mbox{\boldmath$S$}_{1} +  [2+3m_1/(2m_2)]
\mbox{\boldmath$S$}_{2}$. In this sense,  the total energy and
these five independent constants given by Eqs. (7) and (17)
sufficiently determine the integrability of the 10-dimensional
symplectic system. This is another extension to the result of
[8,9] that there is no chaos in the 2PN Hamiltonian formulation of
two compact objects with two spins when the binaries are of equal
mass and spin effects are limited to the leading order spin-orbit
couplings only. Here are two points to emphasize. First, because
the integrability needs only five independent constants, these six
independent constants show that two of five frequencies are
commensurable. In this case, a resonance may occur. Second, it is
not necessary to use the constraint of \emph{equal mass} demanded
in Refs. [8,9], since the existence of these six independent
constants does not depend on any mass. If this constraint is
considered, there seem to be two new additional constants
\begin{equation}
\mbox{\boldmath$S$}\cdot \mbox{\boldmath$S$} = const, \quad \quad
\mbox{\boldmath$L$}\cdot \mbox{\boldmath$S$} = const
\end{equation}
with $\mbox{\boldmath$S$}=\mbox{\boldmath$S$}_1 +
\mbox{\boldmath$S$}_2$. But it should be noted that there are two
relations among Eqs. (7), (17) and (18), $2
\mbox{\boldmath$L$}\cdot \mbox{\boldmath$S$}_{eff} =7
\mbox{\boldmath$L$}\cdot \mbox{\boldmath$S$}$ and
$\mbox{\boldmath$J$}\cdot \mbox{\boldmath$S$} =
\mbox{\boldmath$L$}\cdot \mbox{\boldmath$S$}+
\mbox{\boldmath$S$}^{2}$.  Consequently, no new conserved quantity
appears.

Obviously, the use of the complete symplectic formalism (14) is
rigorous enough to show that the conservative symplectic
Hamiltonian formulations for the two cases of spin effects we
discussed above are completely regular. It makes our procedure
greatly superior to the method of finding parametric solutions to
the Hamiltonian dynamics [8,9] and the technique of fractal basin
boundaries used in [11]. If various higher-order PN expansions are
considered, it is not easy to obtain parametric solutions, and the
fractal method, as a numerical tool for detecting chaos from
order, is a check of the dynamics only in certain situations but
not a full proof of integrability, or nonintegrability. As stated
in [40], ``The only definite proof of integrability is by finding
the analytic forms of the integrals". Our treatment is just fit
for this requirement. In addition, it is worth noting that both
the dynamics of the Hamiltonian formulation and one of the
Lagrangian formulation are completely different in these two spin
effects.

\section{Conclusions}

In the PN Hamiltonian formulation of spinning compact binaries, we
have designed a group of conjugate spin variables that satisfy the
canonical spin-evolution equations. This treatment makes the
complete Hamiltonian formulation symplectic. Seen from Hamiltonian
mechanics and a dynamical system theory, the construction of the
symplectic Hamiltonian formulation is so important that all
properties of Hamiltonian dynamical systems can directly be
applied to the symplectic system. The obtained symplectic
Hamiltonian formulation bring the above-mentioned advantages.
Above all, one of them is that the least number of independent
isolating integrals (equivalent to the half number of all
conjugate state variables) can be regarded to as a criterion for
the prediction of the integrability or the nonintegrability of the
symplectic Hamiltonian binary system. As a result, it is strictly
shown through some theoretical insight that the conservative
symplectic PN Hamiltonian dynamics of spinning compact binaries is
integrable and nonchaotic for the two above-mentioned cases of one
spinning body and the leading spin-orbit interaction.

\begin{acknowledgements}
The authors would like to thank two referees for useful
suggestions. X. W. is supported by the Natural Science Foundation
of China under Contract No. 10873007 and by Science Foundation of
Jiangxi Education Bureau (GJJ09072), and the Program for
Innovative Research Team of Nanchang University. Y. X. is thankful
to the Department of Physics \& Astronomy of the University of
Missouri-Columbia for hospitality and accommodation. Y. X. is
supported by the China Scholarship Council Grant No. 2008102243
and by the National Natural Science Foundation of China (NSFC)
Grant No. 10973009.
\end{acknowledgements}

\end{document}